\documentclass{article}

\usepackage{arxiv}

\usepackage[utf8]{inputenc} 
\usepackage[T1]{fontenc}    
\usepackage{hyperref}       
\usepackage{url}            
\usepackage{booktabs}       
\usepackage{amsfonts}       
\usepackage{nicefrac}       
\usepackage{microtype}      
\usepackage{graphicx}
\usepackage{lipsum}

\title{Complex networks for event detection in heterogeneous high-volume news streams}

\author{
  Iraklis Moutidis \\
  Department of Computer Science\\
  University of Exeter\\
  Exeter, United Kingdom EX4 4QE\\
  \texttt{imoutidi@gmail.com} \\
   \And
  Hywel T.P. Williams \\
  Department of Computer Science\\
  University of Exeter\\
  Exeter, United Kingdom EX4 4QE \\
  \texttt{h.t.p.williams@exeter.ac.uk} \\
}

\begin{document}
\maketitle

\begin{abstract}
Detecting important events in high volume news streams is an important task for a variety of purposes. The volume and rate of online news increases the need for automated event detection methods that can operate in real time. In this paper we develop a network-based approach that makes the working assumption that important news events always involve named entities (such as persons, locations and organizations) that are linked in news articles. Our approach uses natural language processing techniques to detect these entities in a stream of news articles and then creates a time-stamped series of networks in which the detected entities are linked by co-occurrence in articles and sentences. In this prototype, weighted node degree is tracked over time and change-point detection used to locate important events. Potential events are characterized and distinguished using community detection on KeyGraphs that relate named entities and informative noun-phrases from related articles. This methodology already produces promising results and will be extended in future to include a wider variety of complex network analysis techniques.
\end{abstract}


\section{Introduction}
The volume and velocity of online news has increased dramatically in recent years. For news
analysts in various domains (e.g. politics, finance, technology) this creates a need for automated methods to detect and summarize news events in real time, since doing
so with human effort alone is rapidly becoming intractable. News is now consumed online directly from news platforms and aggregators, but also socially via social media, creating a complex media ecosystem. 
Automated methods can assist human analysts by providing alerts to emerging news events, generating brief descriptions of the detected event, and directing the analyst towards relevant documents.
Methods that can be applied to different sources (e.g. online news and social media) are especially useful.

Here we present a methodology that utilizes social network analysis techniques for trending topic detection and characterization in news streams. We make the assumption that news events link multiple entities (e.g. people, organizations, places) at a particular point in time. We hypothesize that news events may therefore be effectively detected by studying the temporal evolution of a knowledge graph that links entities based on their co-occurrence in news documents. In this paper, we explore this approach by developing a software pipeline that creates and analyses complex networks in which the nodes are entities and the edges represent entity co-occurrence in news articles. 
Changes in node degree are used as an indicator of possible news events, which are then characterized using an extended knowledge graph that incorporates noun phrases alongside entities.

To evaluate our method we use two data sets. The first data set consists of articles from the \emph{ALL The News} data set from the Kaggle website\footnote{https://www.kaggle.com/snapcrack/all-the-news}. This data set contains $\approx140,000$ articles from major news sites of the United States of America.
The second dataset is Twitter data relating to the FA Cup Final football match between Chelsea and Liverpool in 2012, which consists of tweets collected during the match by Aiello et al \cite{aiello2013sensing}. We evaluate our method quantitatively and qualitative against a variety of methods tested on that data set in that study.

Our main findings in this paper can be summarized as follows:
\begin{itemize}
\item We outline a named entity method for event detection in news streams, utilising entity co-occurrence relations and change point detection on graph statistics (weighted node degree). 
\item The use of named entities provides an effective way to detect events by tracking a relatively small number of highly informative keywords (compared to n-grams, which typically give a large number of uninformative tokens). This is important for real-time systems which process large volumes of unstructured text.
\item The use of the co-occurrence network provides a novel approach for monitoring each entity, tracking both appearance frequency of an entity and its co-occurrence relationships with other entities.
\item Utilizing KeyGraphs provides a novel approach for event detection and also generates comprehensible summaries of the detected events.
\item Our method out-performs other state of the art techniques for event detection on two types of heterogeneous textual news stream data sets (articles and tweets) for a variety of evaluation metrics.
\end{itemize}

\section{Related Work}
The research area of topic/event detection and tracking (TDT) in news streams has a
long history (e.g. \cite{allan1998topic}, \cite{wayne1997topic}). TDT typically combines natural language processing (e.g. named entity recognition, part of speech tagging, entity disambiguation), information retrieval (e.g. reversed indexing of document keywords, document similarity and clustering), social network analysis (e.g. using relations between
document entities to cluster documents describing the same event into network communities) and machine learning (e.g. extraction of document features to identify context and create clusters).

Event detection research in TDT has two main approaches: document pivot and feature
pivot. The goal of the document pivot approach is to create document clusters that describe
the same event and then extract the appropriate features from them to categorize incoming
articles \cite{allan1998line}.  Petrovic, Osborne, and Lavrenko \cite{petrovic2010streaming} introduce a document pivot approach where Locality Sensitive Hashing is used to cluster documents.
The feature pivot approach focuses on detecting hidden
features to cluster documents and identify news events \cite{he2007analyzing}, \cite{weng2011event}; these approaches are often related to topic models based on Latent Dirichlet Allocation (LDA) \cite{blei2003latent}. 

The authors of \cite{aiello2013sensing} in their BNgram method utilize a $df-idf_t$ score combined with named entity recognition tools to detect and characterize emerging trends on tweets. Another approach to event detection has a focus on word correlations, which are usually measured by distributional similarity \cite{li2003topic} \cite{wartena2008topic} or the number of word co-occurrences \cite{prabowo2008evolving}. Fung, Yu, Yu and Lu \cite{fung2005parameter} worked on detecting important bursty events in text streams. Their technique detects a set of features that correspond with a number of events in a given time window. The features are identified by statistically modelling the frequency of each individual word in an incoming document with a binomial distribution. Then these features are associated with events and time series analysis is used to detect significant changes which could correspond to an important event. This approach uses a large number of features and can be computationally expensive, in addition high frequency words may not be always useful for user interpretation of the content of a detected event. Another approach for event detection with Twitter data is to use recognized named entities in the text, cluster the documents that contain each entity, then apply machine learning algorithms to decide whether the selected documents constitute an event regarding the detected entity \cite{aiello2013sensing}, \cite{popescu2011extracting}, \cite{melvin2017event}.

\section{Method: Network Event Detection (NED) for news streams}
\begin{figure}
\centering
\includegraphics[width=12cm]{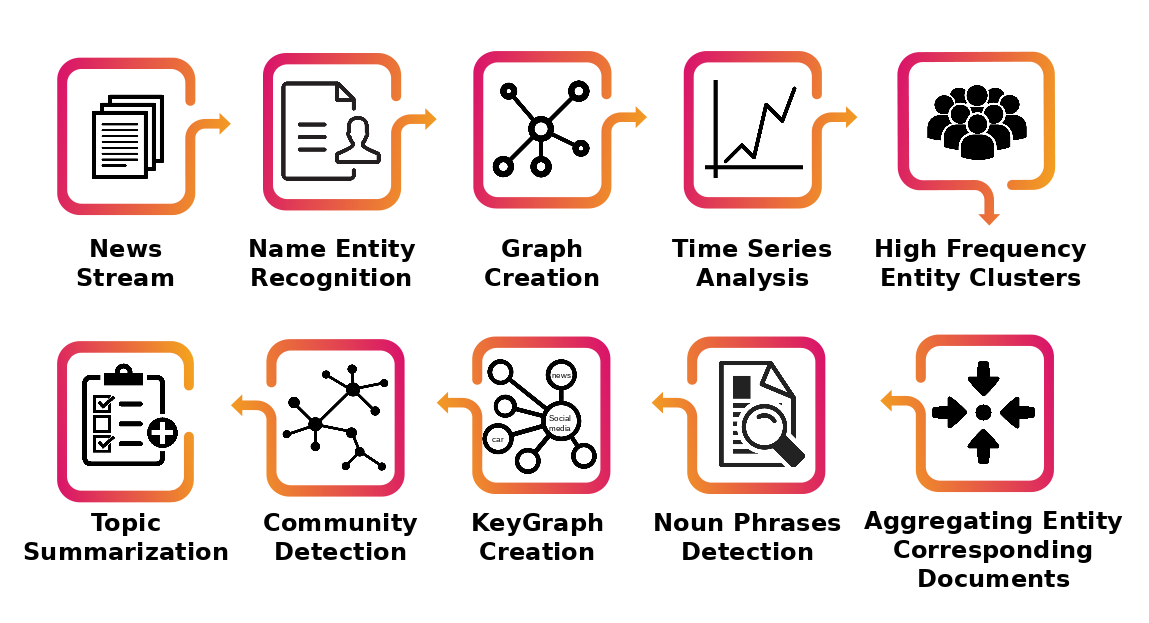}
\caption{Diagram of the main steps of the Named Entity Driven Event Detection (NED) method.}
\label{fig1}
\end{figure}

Here we define a news `event' as a burst of new content related to a particular topic or occurrence; these events may be considered similarly as trending topics. Network Event Detection (NED) makes the assumption that the occurrence of a trending topic or news event (these two terms are used interchangeably hereafter) is signified by changes in prominence of three kinds of named entities: Persons, Locations and Organizations. NED identifies such entities in a stream of documents (articles, tweets) and forms a sequence of networks in which the nodes are unique entities and the edges represent entity co-occurrence within a document. There are two stages to the process: event detection and event characterisation/summarisation. News events are detected by finding peaks in time series associated with individual entities; here we use weighted node degree as the key metric, though other statistics might be tested in future work. The news document stream is filtered down to only retain documents referring to these “peaking” entities, based on a working assumption that these are the most news-worthy entities and that articles that contain them relate to the event. An entity-phrase network is then created to help interpret the detected events, which includes noun phrases as nodes alongside the previously detected entities. Community detection is then used to find communities within the entity-phrase network, with each found community considered to represent a candidate event. Fig.~\ref{fig1} shows the main steps of the process.

\subsection{Entity Detection}
The goal of NED is to detect important events, given a stream of documents (articles, tweets). For each document we apply a named entity recognition (NER) technique in order to detect three kinds of entities: Persons, Locations and Organizations. 
We chose to work with the Stanford NER classifier \cite{finkel2005incorporating} for articles because it is trained with the CoNLL\footnote{http://www.conll.org} dataset which consist of Reuters newswire articles, which is ideal in our case. For detecting named entities in tweets, we chose a different classifier. Tweets are qualitatively different from news articles: they are at most 140 characters long (in the 2012 dataset used here, though the limit has been raised to 280 characters), they contain a lot of misspelled words (typos, jargon, lower case letters), and often lack sufficient context for determining the type of an entity. For entity recognition in tweets we chose the classifier of Ritter et al \cite{ritter2011named}, which according to the authors outperforms the Stanford NER system for this task. 

For news articles, we also supplement entity recognition with a disambiguation process. For Person entities, we replace single words (typically first or last names) with the full name of the entity. Each document is scanned for Person entities. When a single word Person entity is found, it is replaced by the most recent matching multi-word Person entity phrase that was found. If no match is found the single-word name is retained. 
For Location and Organization entities, we disambiguate by expanding abbreviations and reference to manually collated dictionaries of exceptions.

\subsection{Knowledge graph creation}
\label{knowledgeGraph}
\begin{figure}[ht]
\centering
\includegraphics[width=12cm]{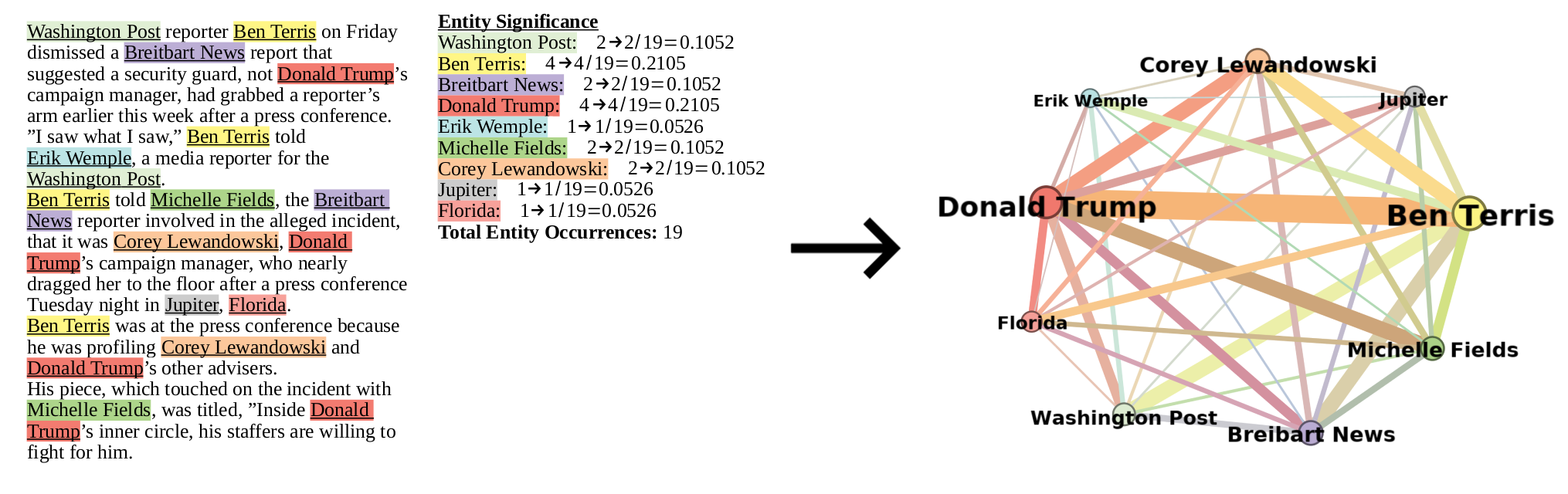}
\caption{Entity significance calculation and the knowledge graph generated from them. The size of the nodes corresponds to their weighted degree and the size of the edges to their weight. The color of each node corresponds to the color of each entity on the left.}
\label{fig3}
\end{figure}

For creating the knowledge graph we use the detected entities as nodes and entity co-occurrence in articles to form weighted edges. In determining edge weight,
we consider the importance of each entity within each article. To do this, we
assign a significance value to each detected entity in a given document as follows:
\begin{equation} \label{eq:1}
S_x(v) = \frac{tf(v,x)} {\sum_{v' \in V} tf(v',x)} 
\end{equation}
where $v$ is the current entity (node), $x$ is a piece of text (article, tweet), $tf(v, x)$ is the term frequency (the raw count of term $v$ in text $x$), and $V$ is the set of all entities in the current document. Fig.~\ref{fig3} shows example output from application of equation \ref{eq:1}. 

The contribution from a document $d$ to the edge weight joining two entity nodes $i, j$ is then given by:
\begin{equation}\label{eq:3}
    w_d(i,j) = \left\{ \begin{array}{cl}
 S_d(i) + S_d(j) 
 &\mbox{ if $i,j \in V$} \\
  0 &\mbox{ otherwise}
       \end{array} \right.
\end{equation}
where $V$ is the set of entities in the current document.
Fig. \ref{fig3} demonstrates the knowledge graph created. 
The aggregation of such graphs from all articles of a given time period forms the overall knowledge graph, according to:

\begin{equation}\label{eq:2}
    W(i,j) = \sum_{d \in D} w_d(i,j)
\end{equation}
where D is the set of all documents.

\begin{figure}[ht]
\centering
\includegraphics[width=11cm]{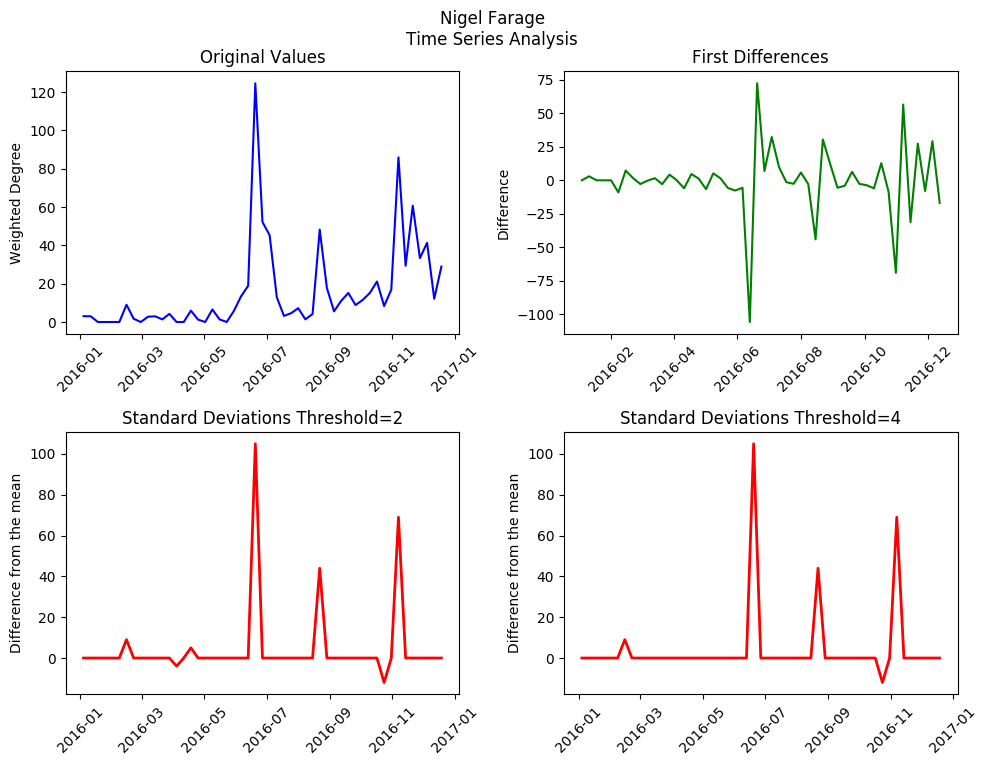}
\caption{Weighted degree time series for the Nigel Farage entity. The first detected peak occurs when the “Brexit” referendum took place, the second highest peak after November occurs when the US Presidential elections took place. The two time series on the bottom, show the significant peaks detected using different thresholds of standard deviations $Y=2$ and $Y=4$.}
\label{time1}
\end{figure}

\subsection{Graph time series analysis}
In the context of this paper we explore detection of news events using as an indicator the significant increase of the weighted degree, the sum of the weights of the adjacent to the node edges, of an entity node. The \emph{All The News} dataset is discretised into 13 one day blocks from 14/6/2016 to 26/6/2016. The FA Cup final data set is discretised into one minute blocks spanning the duration of the game. A sequence of knowledge graphs is created from all blocks in each dataset. Then weighted degree time series are created for all entity nodes from the graph sequence. By monitoring the evolution of weighted degree for all nodes over time, our intention is to detect events based on changes in network structure, revealed in entity time series. 
To detect large changes we calculate the first differences of the time series to remove trends, then we calculate the mean and standard deviation from a sliding window of $X$ blocks. A ``peaking entity'' is then defined as an entity node whose weighted degree exceeds a threshold of $Y$ standard deviations away from the rolling mean. Fig. \ref{time1} illustrates the weighted degree time series analysis we described. In our experiments we used a window of five time blocks ($X=5$) and the threshold distance from the mean was 2 standard deviations ($Y=2$). 

\subsection{Summarizing the detected events}
After identifying all the peaking entities for each time block, event characterization begins by collecting the set of documents for each block that mention the peaking entities. Since multiple events may occur simultaneously, we need to separate out different events in order to gain meaningful information.
This filters out documents that are not relevant to the trending topics we have detected. Doing so gives a substantial improvement in the performance of our method in a quantitative and qualitative manner.

To help distinguish individual events in the same time block we next extract noun phrases from the document set and create a second generation of graphs known as KeyGraphs \cite{sayyadi2009event}.

Simply put, KeyGraphs are the same graphs described in Section~\ref{knowledgeGraph} extended with nouns and noun-phrases, formed only from documents that mention a peaking entity.
For extracting such noun phrases we use the ToPMine algorithm \cite{el2014scalable}.

Next we use the Louvain community detection algorithm \cite{blondel2008fast}, which can be
used on undirected weighted graphs, to identify candidate events as communities in the KeyGraphs.
The entities and noun phrases in each community then form a bag of words summary of the detected event. By sorting this bag of words using the weighted degree of each node we get a comprehensive summary of the event.

\begin{figure}[ht]
\centering
\includegraphics[width=12cm]{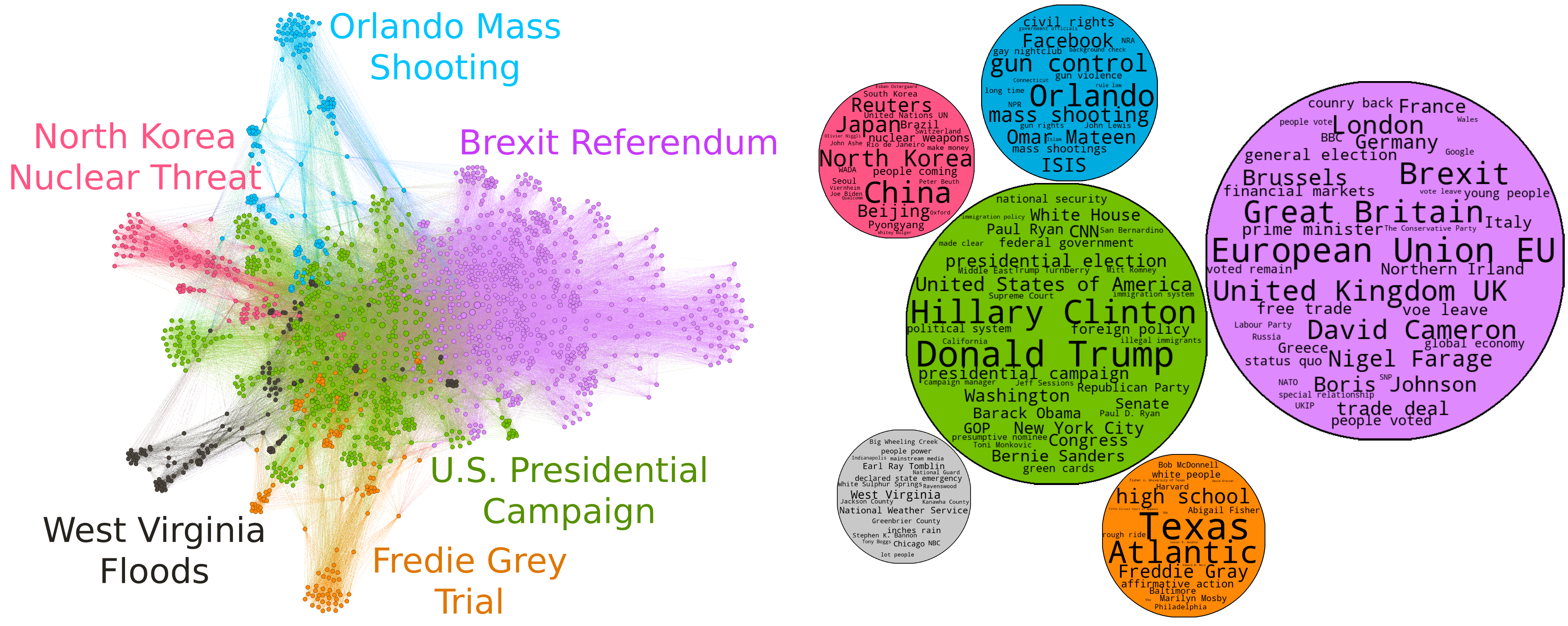}
\caption{Example of KeyGraph method. \emph{Left:} KeyGraph created using the peaking documents of the 24\textsuperscript{th} of June 2016. Community labels are manually added. \emph{Right:} Top entities and noun phrases for each community. The font size corresponds to the weighted degree of each entity/noun phrase node.}
\label{fig4}
\end{figure}

In Fig. \ref{fig4} on the left we present an example KeyGraph generated from documents mentioning peaking entities on the 24\textsuperscript{th} of June 2016 with labels added based on manual inspection of the communities and on the right we present the entities and noun phrases with the highest weighted degree on each community from the same KeyGraph, which together describe the detected event.

\section{Evaluation Method}
To evaluate our method we compare performance against several state of the art techniques.  The reader can refer to Aiello et al \cite{aiello2013sensing} for a description of the first five methods.
The K-Means clustering method \cite{lloyd1982least}, creates clusters of documents, represented by $tf-idf$ vectors which are numeric representations of the text that replace a word from the text with a value based on how often this word appears on the text and how often it appears on other documents. To determine the number of clusters, we ran the algorithm with an initial arbitrary size of 15 clusters and merged similar clusters based on a similarity threshold. Finally, each document cluster corresponds to a detected topic and the terms with the highest $tf-idf$ score were selected to describe it.
The evaluation considers events detected on two datasets, one of newspaper articles and one from Twitter, as described above. 

We used the following event detection methods:
\begin{itemize}
    \item Latent Dirichlet Allocation - LDA 
    \item Document-pivot topic detection - Doc-p
    \item Frequent Pattern Mining - FPM
    \item Soft Frequent Pattern Mining SFPM
    \item BNgram
    \item K-Means clustering
\end{itemize}

\subsection{Evaluation against the FA Cup Final tweet dataset}
This data set consists of tweets that were crawled during the FA Cup Final in 2012, between Chelsea and Liverpool, and was partitioned in one minute time slots. In a previous evaluation of topic detection methods \cite{aiello2013sensing}, the authors of that study reviewed the published media report accounts of the events and chose a set of stories that were significant, time-specific, and well represented on news media to build a topic ground truth. The evaluation test was for each algorithm to recover the topic ground truth. Each ground truth topic consists of a set of \textit{keywords} and a \textit{headline} describing it. For the FA Cup Final data set, a total of 13 one-minute slots were annotated with one topic each (e.g. goals scored, penalties, red/yellow cards). 

We ran our NED method for the FA Cup dataset and calculated the same evaluation metrics that were used in the earlier study \cite{aiello2013sensing}. The aim of this evaluation is for the various methods to generate topics that contain as many keywords as possible from each ground truth topic. Each method returns the top 2 detected topics, each one of them described by a set of n-grams, which in our case  is a contiguous sequence of one or more words from a given sample of text.

Three metrics were used to evaluate the topic detection methods on this data set:
\begin{itemize}
\item \textit{Topic recall}: Percentage of ground truth topics successfully detected by a method. A ground truth topic was considered to be successfully detected when an algorithm produced a topic with a set of keywords that contained all mandatory keywords in the ground truth topic. 

\begin{equation} \label{eq:T_Recal}
Topic\ Recall = \frac{Relevant\ Topics\ Detected\ by\ the\ Method} {All\ Relevant\ Dataset\ Topics} 
\end{equation}

\item \textit{Keyword precision}: Percentage of correctly detected keywords (as described above) from the total number of keywords for the detected topics matched to some ground truth topic in the time slot under consideration. The total precision of a method is computed by micro-averaging the individual precision scores over all time slots.

\begin{equation} \label{eq:K_Precision}
Keyword\ Precision = \frac{Relevant\ Keywords\ Detected\ by\ the\ Method} {All\ Keywords\ Detected\ by\ the\ method} 
\end{equation}

\item \textit{Keyword recall}: Percentage of correctly detected keywords from the total number of keywords for the ground truth topics that have been matched to some detected topic in the time slot under consideration. The total recall is similarly computed by micro-averaging.

\begin{equation} \label{eq:K_Recall}
Keyword\ Recall = \frac{Relevant\ Keywords\ Detected\ by\ the\ Method} {All\ Relevant\ Dataset\ Keywords} 
\end{equation}

\end{itemize}

\subsection{Evaluation against the All The News data set}
To evaluate the different methods on the articles of the ALL The News data set we first created a ground truth set of events by manually annotating 1861 articles from an eight-day period from 19/6/2016 until 26/6/2016. For each article, we annotated the main topics and then created groups of articles based on the annotated topics. We also used articles from five previous days for the methods that required a time window of previous days to detect emerging trends (NED, BNgram). The time slot for this data set was one day. Multiple articles that discuss the same subject were considered to form a topic; for each day, the topics that were discussed by more than 4 articles were taken as the ground truth topics to be detected in the evaluation. 

For the \emph{All the News} evaluation, we used implementations of the LDA, Doc-p and BNgram methods from \cite{aiello2013sensing} and we also implemented the K-Means approach. For all the methods, we manually inspected which of the returned groups of n-grams (detected topics) matched the annotated ground truth topics. An example of the generated groups of n-grams generated by each method can be found in Table \ref{descr}.

To evaluate the topic detection methods we used two metrics:
\begin{itemize}
\item \textit{Precision}: Percentage of ground truth topics successfully detected by each method within the total number of candidate topics returned.

\begin{equation} \label{eq:T_Precision}
Precision = \frac{Relevant\ Topics\ Detected\ by\ the\ Method} {All\ Detected\ Topics\ by\ the\ Method} 
\end{equation}

\item \textit{Recall}: Percentage of ground truth topics successfully detected by each method within the total number of ground truth topics and it is equivalent with equation \ref{eq:T_Recal} .
\end{itemize}

\section{Evaluation Results}
\label{results}
Our proposed NED algorithm outperformed all of the comparator methods on most of the metrics that were calculated in both evaluations. Results for the \emph{All The News} evaluation are presented in Table \ref{atnr} and results for the FA Cup Final evaluation are presented in Table \ref{facf}.

\begin{table}[t] 
\caption{FA Cup Final data set evaluation results}
\centering 
\footnotesize
\begin{tabular*}{7cm}{l @{\hskip 1cm} c  c c} 
\textbf{Method} &  T-REC & K-PREC & K-REC\\ 
\midrule
\midrule
LDA  & 69.23\% & 16.37\% & 68.29\% \\
Doc-p & 76.92\% & 33.73\% & 58.33\% \\
FPM & 30.77\% & \textbf{75\%} & 42.86\% \\
SFPM & 61.54\% & 23.36\% & 65.79\% \\
BNgram & 76.92\% & 29.89\% & 57.78\% \\
NED  & \textbf{84.61}\% & 24.74\% & \textbf{79.31}\% \\
\midrule 
\bottomrule 
\end{tabular*}
\label{facf} 
\end{table}

\begin{table}[t]
\caption{All The News data set evaluation results}
\centering 
\footnotesize
\begin{tabular*}{7cm}{l@{\hskip 1.4cm} c @{\hskip 1.4cm}c} 
\textbf{Method} & Precision & Recall\\ 
\midrule
\midrule
LDA  & 40.8\% & 40.8\% \\
Doc-p & 40.7\% & 21.9\% \\
BNgram & 53.95\% & 38.04\% \\
K-Means & 51.24\% & 35.32\% \\
NED   & \textbf{66.98}\% & \textbf{50.48}\% \\
\midrule 
\bottomrule 
\end{tabular*}
\label{atnr} 
\end{table}

For the FA Cup Final dataset which consists of Twitter posts, the NED method outperformed the rest of the methods in topic recall and keyword recall. NED did not perform as well as three of the methods (FPM, Doc-p and BNgram) in keyword precision. The reason for this is that our method produces a relatively large number of keywords to describe each topic and for that reason the keyword precision drops significantly. Restricting the number of keywords describing each topic can efficiently address this issue (data not shown) but it has an adverse effect on topic recall and keyword recall. 

For the the \emph{All The News} data set which consists
of longer news articles, NED achieved substantially better precision and recall than all the other methods tested.

\section{Event Summarisation}

Although it is hard to quantify the quality of a given summary of an event or trending topic, we present some examples from the \emph{All The News} evaluation that we believe qualitatively demonstrate the good performance of the NED method on this aspect of the event detection task. Table \ref{descr} gives the top n-grams generated by each method for the detected event corresponding to the Brexit referendum that occurred on 24\textsuperscript{th} June 2016, one day after the referendum took place (note that news articles are typically published the day after the event takes place). 

\begin{table}[t]
\caption{Brexit Referendum topic descriptions}
\centering 
\footnotesize
\begin{tabular}{|l|l|}
\hline
\textbf{Method} & \multicolumn{1}{c|}{\textbf{Topic Description}}                                                                                                               \\ \hline
LDA             & \begin{tabular}[c]{@{}l@{}}European, brexit, Britain, new, Trump, vote, EU, political, leave,\\ it’s going, Clinton, UK, campaign, May, united, president, two,\\ British,  first, economic\end{tabular}                                                                                                                         \\ \hline
Doc-p           & \begin{tabular}[c]{@{}l@{}}independence, remain, interdependent, heralded, British,\\ countervailing, brexit, institutionalists, Farage, proximate, pen’s,\\ EU, globalist, xenophobes, pollsters, unbridled, Hillary,  globalism,\\ foreshadowing, comeuppance\end{tabular}                                                       \\ \hline
BNgram          & \begin{tabular}[c]{@{}l@{}}people, Brexit, vote, European Union, two, United Kingdom, new,\\ even, referendum, last, get, told, years, want, time, leave, political,\\  British, May, back\end{tabular}                                                                                                                          \\ \hline
K-Means         & \begin{tabular}[c]{@{}l@{}}EU, Britain, European, brexit, trump, leave, vote, British, union,\\ UK, Europe, United, referendum, voters, remain, economic,\\ Cameron,  Kingdom, trade, Farage\end{tabular}                                                                                                                         \\ \hline
NED           & \begin{tabular}[c]{@{}l@{}}European Union EU, Great Britain, United Kingdom UK, London,\\ Brexit, David Cameron, Nigel Farage, Brussels, Boris Johnson,\\  Germany, trade deal, prime minister, Northern Ireland,\\ general election, vote leave, country back, special relationship,\\ people vote, UKIP, vote leave\end{tabular} \\ \hline

\end{tabular}
\label{descr}
\end{table}

\section{Discussion}
This paper presents a prototype news event detection methodology, based on natural language processing and network analysis. Events are located by finding peaks in the prominence of named entities, based on node-level time series in the entity knowledge graph and characterized by community detection in KeyGraphs linking entities and noun-phrases. 
Our evaluation suggests that NED provides a significant improvement against other state of the art methods. This is supported on two qualitatively different datasets (news articles and tweets). The combination of named entities and social network analysis techniques, such as community detection, seems to be very effective in detecting and tracking topics in document streams, and provides a more comprehensive description of each detected event, compared with the rest of the evaluated methods. 

We suggest that NED performs better than other methods for detecting events in document news streams because it focuses on named entities (highly relevant to news events) and, in particular, it identifies `peaking entities' which show an increased level of prominence within the dynamic knowledge graph. This helps to remove irrelevant articles from further processing; here such articles constitute `noise' in the document stream and do not contribute anything to the detection and description of the trending topics or events. The approach removes a substantive proportion of articles in this way, in some cases  around 50\% of them. Another advantage of our method is the use of named entities and noun phrases to form communities of n-grams that describe an event. 

By utilizing the entity co-appearance network, the NED method captures relational information about entity interactions. Thereby it can efficiently detect peaking entities based on an approach where an entity is important not only because it appears frequently itself, but also because it co-occurs with other important entities. The use of relational information allows event detection to access holistic patterns in articles. While here we have focused on weighted degree as a node-level indicator of structural change, future work will consider other network statistics including macro-/meso-level attributes such as community structure, core-periphery, and backbone topology.

\subsubsection*{Acknowledgements}
The authors acknowledge funding from a commercial entity, Adarga Ltd (https://www.adarga.ai). The funder had no input or editorial influence over the manuscript.

\bibliographystyle{unsrt}  


\end{document}